\documentclass[review]{cas-sc}

\usepackage[authoryear]{natbib}
\usepackage{mathrsfs}
\usepackage{multirow}
\usepackage[normalem]{ulem}
\useunder{\uline}{\ul}{}

\def\tsc#1{\csdef{#1}{\textsc{\lowercase{#1}}\xspace}}
\tsc{WGM}
\tsc{QE}
\tsc{EP}
\tsc{PMS}
\tsc{BEC}
\tsc{DE}

\begin{document}
\let\WriteBookmarks\relax
\def\floatpagepagefraction{1}
\def\textpagefraction{.001}

\shorttitle{Automatic parotid tumor segmentation}

\shortauthors{Yifan Gao et~al.}

\title [mode = title]{An Anatomy-aware Framework for Automatic Segmentation of Parotid Tumor from Multimodal MRI}                      


%

\author[1,2]{Yifan Gao}[type=editor,
						orcid=0000-0002-9184-0085,
						style=chinese]
\fnmark[1]

\ead{yifangao@mail.ustc.edu.cn}

\credit{Conceptualization of this study, Methodology, Software}

\address[1]{College of Medicine and Biological Information Engineering, Northeastern University, Shenyang, 110169, China}

\address[2]{School of Biomedical Engineering (Suzhou), Division of Life Sciences and Medicine, University of Science and Technology of China, Hefei, 230026, China}

\address[3]{Engineering Center on Medical Imaging and Intelligent Analysis, Ministry Education, Northeastern University, Shenyang, 110169, China}

\address[4]{Department of Oromaxillofacial-Head and Neck Surgery, School of Stomatology, China Medical University, Shenyang, 110002, China}

\address[5]{Liaoning Jiayin Medical Technology Co., LTD, Shenyang, 110170, China}

\author[1,3]{Yin Dai}[style=chinese]
\fnmark[1]
\cormark[1]

\ead{daiyin@bmie.neu.edu.cn}

\author[4]{Fayu Liu}[style=chinese]
\credit{Data curation, Writing - Original draft preparation}

\author[1,3]{Weibing Chen}[style=chinese]

\author[5]{Lifu Shi}[style=chinese]

\fntext[1]{These authors contributed equally to this work.}
\cortext[cor1]{Corresponding author}

\begin{abstract}
Magnetic Resonance Imaging (MRI) plays an important role in diagnosing the parotid tumor, where accurate segmentation of tumors is highly desired for determining appropriate treatment plans and avoiding unnecessary surgery. However, the task remains nontrivial and challenging due to ambiguous boundaries and various sizes of the tumor, as well as the presence of a large number of anatomical structures around the parotid gland that are similar to the tumor. To overcome these problems, we propose a novel anatomy-aware framework for automatic segmentation of parotid tumors from multimodal MRI. First, a Transformer-based multimodal fusion network PT-Net is proposed in this paper. The encoder of PT-Net extracts and fuses contextual information from three modalities of MRI from coarse to fine, to obtain cross-modality and multi-scale tumor information. The decoder stacks the feature maps of different modalities and calibrates the multimodal information using the channel attention mechanism. Second, considering that the segmentation model is prone to be disturbed by similar anatomical structures and make wrong predictions, we design anatomy-aware loss. By calculating the distance between the activation regions of the prediction segmentation and the ground truth, our loss function forces the model to distinguish similar anatomical structures with the tumor and make correct predictions. Extensive experiments with MRI scans of the parotid tumor showed that our PT-Net achieved higher segmentation accuracy than existing networks. The anatomy-aware loss outperformed state-of-the-art loss functions for parotid tumor segmentation. Our framework can potentially improve the quality of preoperative diagnosis and surgery planning of parotid tumors.
\end{abstract}



\begin{keywords}
Parotid tumor segmentation \sep Multimodal fusion \sep Anatomy-aware loss \sep Deep learning \sep Transformer 
\end{keywords}

\maketitle

\section{Introduction}

Parotid tumors are the most common salivary gland tumors, accounting for approximately 2\% to 6\% of head and neck tumors \cite{jones2008range}. Parotid tumors are categorized into five types according to their clinical characteristics: pleomorphic adenomas, Warthin tumors, basal cell adenomas, malignant tumors, and other minor benign lesions \cite{jang2004basal,mendenhall2008salivary,zheng2021development}. It has been estimated that about 20\% of all parotid tumor cases are malignant \cite{bussu2011clinical}.

Despite the relatively low incidence of the parotid tumor, the rate of clinical misdiagnosis before surgery is high due to the heterogeneity and the diversity of types \cite{assadsangabi2022common}. Currently, the primary treatment for parotid tumors is resection surgery \cite{poletti2018multiplanar}. Inappropriate surgical planning, however, can result in incomplete tumor resection or damage to the facial nerves \cite{espinosa2018clinicopathologic,stathopoulos2018partial,grasso2021rupture}. On the one hand, incomplete tumor resection may lead to the recurrence of the tumor. Even more dangerously, re-operation to remove the tumor can be a very complicated process, and there is an unfavorable prognosis for the patient \cite{abu2016recurrent,kanatas2018current}. On the other hand, loss of facial nerve function in severe cases can lead to permanent facial paralysis, significantly impacting the patient's postoperative recovery and quality of life \cite{tseng2007malignant}. Therefore, developing techniques for accurate and personalized preoperative diagnosis of parotid tumor patients is of crucial importance \cite{matsuo2020diagnostic,dai2021transmed,dai2022mutual}.

Modern medical imaging plays an essential role in the preoperative diagnosis of parotid tumors. Among them, multimodal Magnetic Resonance Imaging (MRI) can provide the most accurate results for diagnosing tumors due to its good contrast and rich information \cite{soler1997pictorial,stoia2021cross}. For preoperative diagnosis and quantitative assessment of the disease, automatic parotid tumor segmentation from MRI is necessary. As performing manual segmentation of the parotid tumors from 3D volumes is tedious, time-consuming, and often interferences by anatomical structures within or around the parotid gland, automatic segmentation of tumors is highly preferable in clinical practice.

Recent advances in artificial intelligence, notably deep learning, have contributed to significant breakthroughs in image recognition and have been widely applied in data-driven medical image analysis. Deep learning models depicted by convolutional neural networks (CNNs) \cite{krizhevsky2017imagenet,he2016deep} and fast-evolving Transformer \cite{vaswani2017attention,dosovitskiy2020image,liu2021swin} have achieved impressive performance in various medical image tasks, such as computer-aided diagnosis \cite{jian2021multiple,jian2022mri,zhao20213d,chen2022gashis,amador2022hybrid} and medical image segmentation \cite{zhao2020deep,wang2020automatic,wang2021semi,elsawy2022pipe}. These advances have demonstrated that deep learning-based technologies are promising for studying the automatic segmentation of parotid tumors in multimodal MRI.

\begin{figure}
	\centering
	\includegraphics[scale=.8]{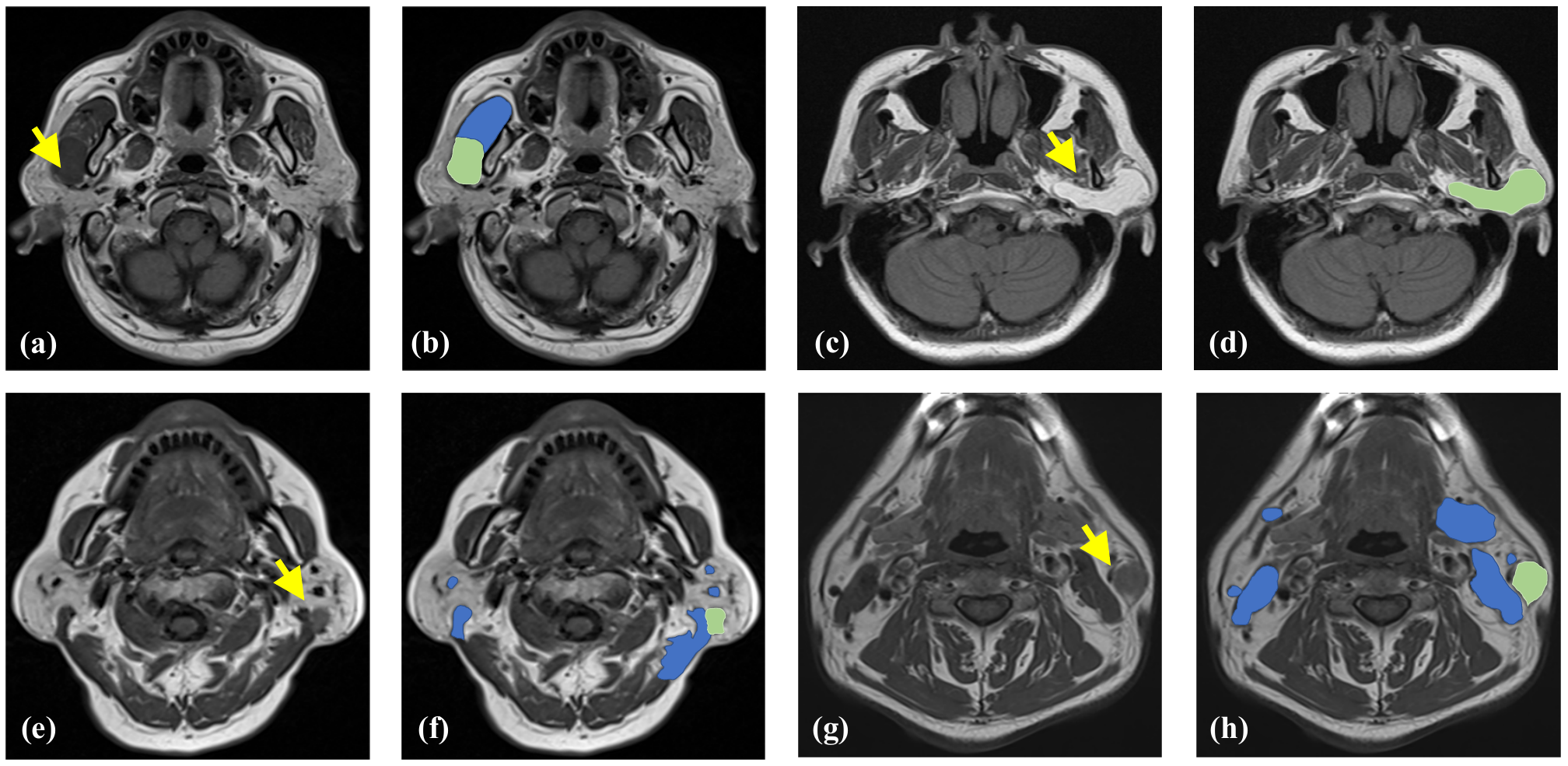}
	\caption{Schematic diagram of parotid MRI. Figures (a), (c), (e), and (g) indicate the original image, while (b), (d), (f), and (h) show the ground truth corresponding to it. The yellow arrow denotes the location of the parotid tumor. The green mask indicates the actual segmentation label, and the blue mask means the position where the deep model is likely to make mistakes.}
	\label{FIG:1}
\end{figure}

Despite the success of state-of-the-art segmentation techniques in multiple organs and lesions, it has not been applied to the parotid tumor to provide personalized treatment. In addition to the limited clinically available imaging data, there are two important challenges to overcome.

On the one hand, the parotid gland and its surroundings are complex, with a large number of anatomical structures. This characteristic makes it challenging to develop automatic segmentation techniques. The major anatomical structures within and around the parotid gland can be roughly divided into muscular tissues and neurovascular structures. The muscular tissues include the internal medial pterygoid muscle, the posterior belly of digastric muscle, and the sternocleidomastoid muscle. The neurovascular structures mainly include the retromandibular vein, internal carotid artery, internal jugular vein, external carotid artery, and other facial nerves. In most cases, the facial nerve is not discernible on MRI \cite{prevost2022external}. As other vessels are usually visible on MRI, they may also undergo anatomical variation in the case of tumor-occupying lesions and not be visible \cite{prevost2022external}.

These anatomical structures often have similar signal intensities to the tumor in MRI and even adhere together, bringing major problems to automatic segmentation. Fig. \ref{FIG:1} demonstrates some hard samples in parotid tumor segmentation. The comparison in Fig. \ref{FIG:1}(a) and \ref{FIG:1}(b) shows that the tumor and the muscle tissue have almost the same signal intensity. It is difficult to differentiate the parotid tumor from this anatomical structure in this case. In addition, the scale of parotid tumors is highly variable, ranging from less than one millimeter to several centimeters in radius. In the small tumor segmentation, the model is highly likely to confuse it with vascular tissues or muscle tissues. The comparisons from Fig. \ref{FIG:1}(e) to 1(h) highlight many anatomical structures with similar intensity and shape to the parotid tumor. Therefore, this feature makes it very hard for the model to focus on the segmentation of the ground truth tumor. In addition, the location of the parotid gland in the images often shows a small amount of signal from the facial nerve. As seen in Fig. \ref{FIG:1}(e), it further increases the difficulty of automatic segmentation. In summary, unlike most organs and lesions, the automatic segmentation of parotid tumors is a challenging task. It requires the introduction of prior anatomical knowledge to improve the robustness and reliability of the model.

On the other hand, parotid tumors themselves have a large number of types. The signal intensity, morphology, and size of different tumor types in MRI are very different. Therefore, it is difficult for the deep learning-based model to learn robust feature information related to tumors. As seen in Fig. \ref{FIG:1}(c) and \ref{FIG:1}(d), tumors exhibit higher signal intensities, making it difficult to distinguish them from parotid glands and compare and analyze them with other tumors. 

However, experienced radiologists have good consistency in parotid tumor segmentation. The critical factor in this is the effective extraction and combination of multimodal image information by the expert, which allows for accurate manual segmentation. The parotid MRI examination produces multiple images. Among them, the most informative and commonly used modalities are T1 images, T2 images, and STIR images. Although anatomical variations are very common in individual MRI modalities, it is rare for parotid anatomy and tumors to have abnormal morphology and signal intensity in all three  slices. Therefore, the expert can consider and make the final decision based on observation and comparison of the different image slices in a comprehensive manner. In summary, deep models for parotid tumor segmentation need to learn cross-modal representations from multimodal MRI and fuse features from three modalities to improve the model's performance.

Therefore, this paper develops an anatomy-aware framework for automatic segmentation of parotid tumors from multimodal MRI to leverage rich anatomical prior knowledge. The framework contains the Transformer-based segmentation network (PT-Net) and the anatomy-aware loss function.

First, we propose PT-Net, a novel Transformer-based coarse-to-fine multimodal fusion network for parotid tumor segmentation. The encoder of the network is built on the Transformer, while the decoder is CNNs-based architecture. Such a design has been shown to balance local feature extraction and global information modeling. Different from the existing multimodal fusion approaches, the encoder extracts and merges contextual information from three modality-specific parotid MRI at different scales. It can better obtain cross-modality and multi-scale tumor information. The decoder stacks the feature maps of various modalities and calibrates the multimodal information using the channel attention mechanism. Experiments demonstrate that our method has significant advantages over highly competitive baseline methods in parotid tumor segmentation.

Second, this paper presents the anatomy-aware loss for guiding the deep model to distinguish parotid anatomical structures from tumors. Considering that segmentation models are prone to be disturbed by irrelevant anatomy and make wrong predictions, we develop this novel distance-based loss function. In contrast to previous methods \cite{kervadec2019boundary,karimi2019reducing}, anatomy-aware loss uses the distance of the center coordinates for computing the binary mask in both model-predicted segmentation and ground truth. Hence this loss function can force the model to identify anatomical structures far from the ground truth, thereby predicting the correct tumor location. It is worth noting that compared with other distance-based loss functions applied to medical image segmentation, our anatomy-aware loss does not require additional computation and has high training stability.

Based on our experimental results with MRI of 187 parotid tumor patients, we demonstrated the effectiveness of the proposed PT-Net and the anatomy-aware loss. This approach has the potential to reduce the annotation burden associated with large-scale parotid tumor image datasets, as well as mitigate the difficult availability of high-quality labels provided by experienced radiologists. 

The main contribution of this paper is summarized as follows:

\begin{enumerate}
	\item We study parotid tumor segmentation for the first time and propose an automatic segmentation framework with high performance and robustness.
	\item We propose a Transformer-based segmentation network that fuses multimodal information from coarse to fine. The proposed PT-Net captures compact and high-level tumor features through the self-attention mechanism.
	\item This work presents the anatomy-aware loss function. It exploits the prior knowledge of anatomy in parotid MRI to reduce segmentation mistakes from tumor-similar anatomical structures.
\end{enumerate}

The remaining of this paper is organized as follows: In Section 2, we describe the proposed automatic segmentation framework in detail, including the multimodal fusion network called PT-Net, and the anatomy-aware loss function. In Section 3, we show the experiment design and the results. The experiment results are further analyzed and discussed in Section 4. Finally, we summarize our work in Section 5.

\section{Method}

\subsection{Parotid tumor segmentation network}

\begin{figure}
	\centering
	\includegraphics[scale=.65]{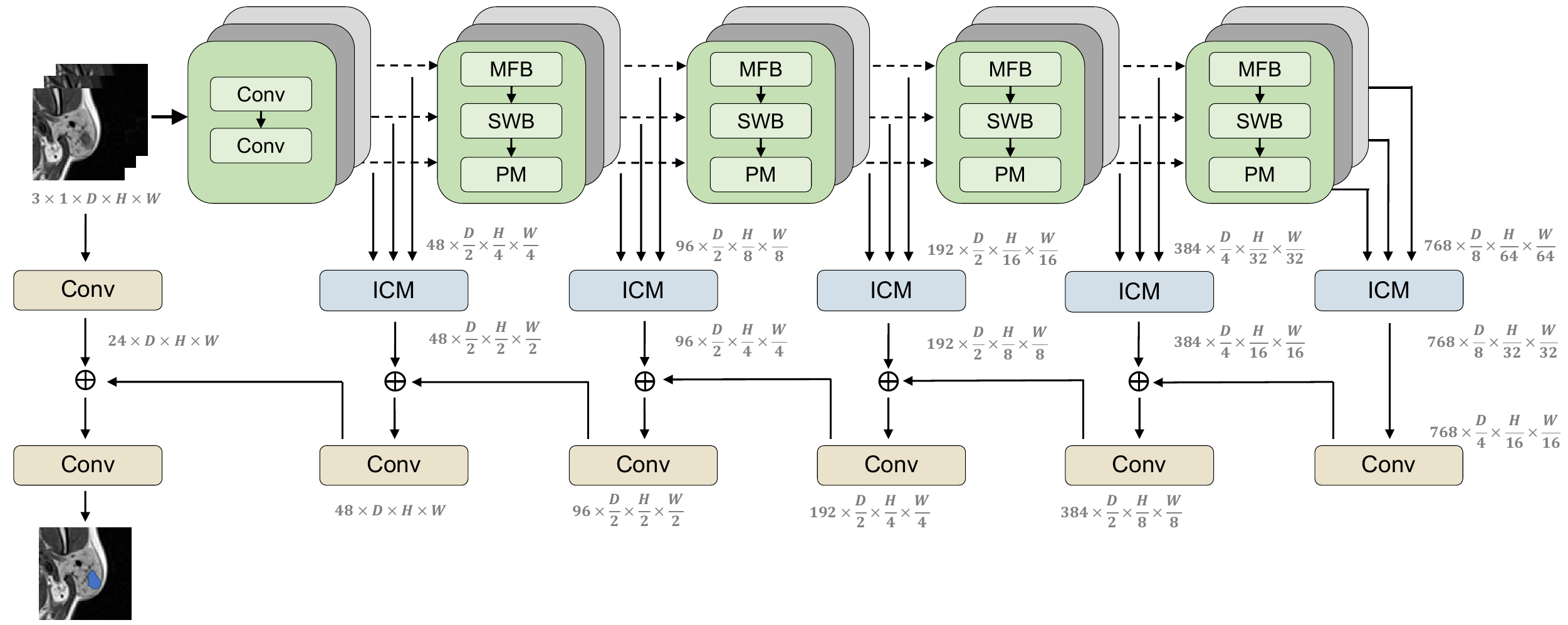}
	\caption{Overview of the proposed PT-Net, which consists of the Transformer-based encoder and CNNs-based decoder. Conv: Convolution Block. MFB: Multimodal Fusion Block. SWB: Shift Window Block. PM: Patch Merging. ICM: Information Calibration Module.}
	\label{FIG:2}
\end{figure}

\subsubsection{Network architecture overview}
PT-Net follows the classical design of encoder-decoder architecture \cite{ronneberger2015u}, which uses both CNNs and Transformer for modeling low-level features and long-range dependencies. Compared with other Transformer-based segmentation networks, our method uses a novel multimodal fusion module. The proposed approach can effectively extract and fuse information from coarse to fine to improve the segmentation performance of the model. Fig. \ref{FIG:2} shows the basic structure of our proposed model. In the next section, we will elaborate on the structure of the segmentation network, which is the Transformer-based encoder and CNNs-based decoder, respectively. Besides, we will also introduce the multimodal fusion block, an important component of our PT-Net.

\subsubsection{Transformer-based encoder}
As seen in Fig. \ref{FIG:2}, the encoder of PT-Net consists of three modality-independent networks that do not share parameters. Assume that the input of each modality is a randomly cropped 3D patch $X \in R^{1 \times D \times H \times W}$ from the original image, where $D$, $H$, and $W$ denote the depth, height, and width of each image, respectively.

First, similar to \cite{zhou2021nnformer}, the image of each modality is fed to the embedding layer to reduce the original resolution. The embedding layer contains two successive convolutional blocks. Each convolution block contains two convolutional layers with a kernel size of 3. After each convolution layer, we perform the GELU activation function \cite{hendrycks2016gaussian} and instance normalization \cite{ulyanov2016instance}. Since the 3D parotid MRI data used in this paper is significantly smaller in depth than the other two dimensions, the downsampling ratios of the three dimensions are different in the two convolution layers. Specifically, we reduce the stride of the depth dimension of the first convolution block from 2 to 1. Therefore, the downsampling ratio is 2 for the depth dimension and 4 for the other two dimensions. Finally, we obtain the high-dimensional vector $X_{e} \in R^{C \times \frac{D}{2} \times \frac{H}{4} \times \frac{W}{4}}$, and $C$ is the length of the representation vector. We summarize the specific structure of the embedding layer in Table \ref{tbl1}.

\begin{table}[width=.7\linewidth,cols=4,pos=h]
	\centering
	\caption{Detailed configuration of the embedding layer. Note that the default input patch size is 8×320×320.}
	\label{tbl1}
	\begin{tabular*}{\tblwidth}{ccccc}
		\hline
		Layer                              & Block  & Stride    & Input shape       & Output shape      \\ \hline
		\multirow{2}{*}{Embedding layer 1} & Conv3d & [2, 2, 2] & [1, 8, 320, 320]  & [24, 4, 160, 160] \\
		& Conv3d & [1, 1, 1] & [24, 4, 160, 160] & [24, 4, 160, 160] \\ \hline
		\multirow{2}{*}{Embedding layer 2} & Conv3d & [1, 2, 2] & [24, 4, 160, 160] & [48, 4, 80, 80]   \\
		& Conv3d & [1, 1, 1] & [48, 4, 80, 80]   & [48, 4, 80, 80]   \\ \hline
	\end{tabular*}
\end{table}

Next, the feature map of the three modalities is fed into four stages of structurally identical Transformer blocks. Each Transformer block contains a multimodal fusion block, a shift-window block, and a patch merging operation. As shown in Fig. \ref{FIG:2}, given the input feature maps of the three modalities as $X_{1}$, $X_{2}$, and $X_{3}$, the computation procedure of the encoder at stage $l$ can be summarized as follows:

\begin{equation}
	\begin{gathered}
		\hat{X}_{1}^{l} = \textit{MFB}(X_{1}^{l-1}, X_{2}^{l-1}, X_{3}^{l-1}) \\
		X_{1}^{l} = \textit{PM}(\textit{SWB}(\hat{X}_{1}^{l}))
	\end{gathered}
\end{equation}

MFB is the abbreviation of multimodal fusion block, SWB denotes shift-window block, and PM means patch merging. $\hat{X}_{1}^{l}$ and $X_{1}^{l}$ donate the output of the multimodal fusion block and the output of the patch merging module in stage $l$, respectively.

Unlike other medical image fusion methods, we designed a novel multimodal fusion block. It establishes a cross-modality information fusion path from coarse to fine and enables feature communication and flow under multi-scale and multi-location conditions. We elaborate on the implementation of the multimodal fusion block in the next section. The implementation details of the sliding window block and patch merging operations are the same as those of the state-of-the-art computer vision backbone Swin Transformer \cite{liu2021swin}. Given an input feature map $X_{1}$, the sliding window block for the $l$-th stage is implemented in the following formulation:

\begin{equation}
	\begin{gathered}
		\hat{Z}_{1}^{l} = \textit{SW-MSA}(\textit{LN}(\hat{X}_{1}^{l})) + \hat{X}_{1}^{l} \\
		Z_{1}^{l} = \textit{MLP}(\textit{LN}(\hat{Z}_{1}^{l})) + \hat{Z}_{1}^{l}
	\end{gathered}
\end{equation}

MLP means multilayer perceptron, SW-MSA stands for sliding window multi-head self-attention module, and LN is layer normalization. $\hat{Z}_{1}^{l}$ and $Z_{1}^{l}$ denote the output features from SW-MSA and MLP, respectively. Sliding windows can effectively capture the location of lesions in the feature map at multiple scales and locations, thus facilitating the model to extract high-level representations dynamically. Finally, the feature map is fed to the patch merging module. The merging operation consists of the GELU activation function and layer normalization, and the resolution becomes half of the original one by 3D convolution with a stride of 2. As shown in Fig. \ref{FIG:2} the first and second stages reduce the size of the feature maps only in the length and width dimensions, considering the low resolution of depth in parotid tumor MRI. The patch merging module can be presented as follows:

\begin{equation}
	\begin{aligned}
		\hat{X}_{1}^{l+1} = F_c(\textit{LN}(\textit{GELU}(Z_{1}^{l})))
	\end{aligned}
\end{equation}

where $F_c$ denotes the 3D convolution function.

\begin{figure}
	\centering
	\includegraphics[scale=.7]{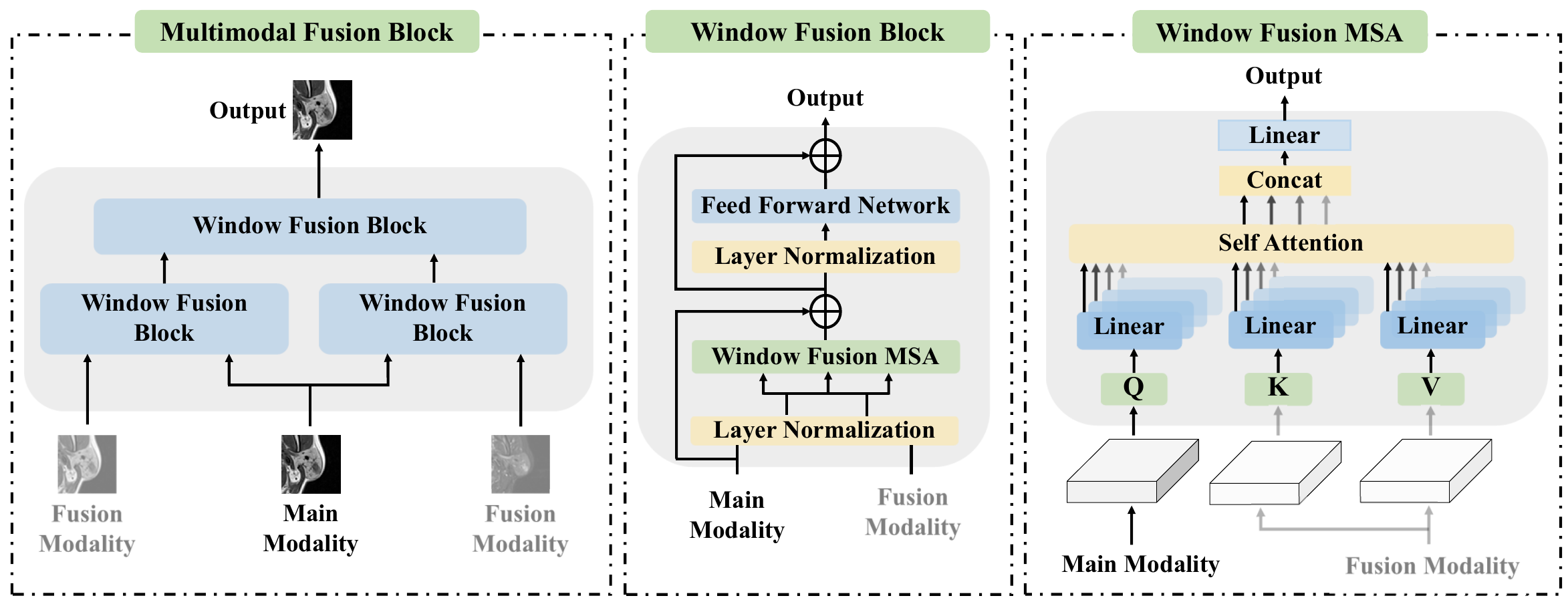}
	\caption{Detailed structure of our proposed multimodal fusion block.}
	\label{FIG:3}
\end{figure}

\subsubsection{Multimodal fusion block}
Fig. \ref{FIG:3} shows the detailed architecture of the multimodal fusion block in PT-Net. Here we describe the design and implementation of our proposed block. As can be seen from the schematic diagram, the input of each modally independent encoder consists of one modality of the major branch and two modalities of the minor branch. As shown in Fig. \ref{FIG:3}(a), the multimodal fusion block consists of three window fusion blocks. The feature map of the main modality is input to the two window fusion blocks as the original information. In contrast, the feature maps of the two minor modalities are each input to one of the window fusion blocks as the fusion information. Given the input vector $X_1$ for the main modality and the input vectors $X_2$ and $X_3$ for the fusion modality, the multimodal fusion block in stage $l$ can be represented as follows:

\begin{equation}
	\begin{gathered}
		X_{12}^{l} = \textit{WFB}(X_{1}^{l-1}, X_{2}^{l-1}) \\
		X_{13}^{l} = \textit{WFB}(X_{1}^{l-1}, X_{3}^{l-1}) \\
		\hat{X}_{1}^{l} = \textit{WFB}(X_{12}^{l}, X_{13}^{l})
	\end{gathered}
\end{equation}

WFB stands for window fusion block, $X_{12}^{l}$ and $X_{13}^{l}$ are the outputs of the two fusion blocks at the bottom of Fig. \ref{FIG:3}(a). The implementation of the window fusion block is similar to \cite{liu2021swin}. However, with one significant modification, as shown in Fig. \ref{FIG:3}(b). The self-attention matrix is no longer obtained from a single feature map by linear projection. Instead, we use the feature maps of the major and minor modalities as inputs. Given the input vectors $X_{1}$ for the major modality and $X_{2}$ for the minor modality, the window fusion block for stage $l$ can be expressed as follows:

\begin{equation}
	\begin{gathered}
		\hat{X}_{12}^{l} = \textit{WF-MSA}(\textit{LN}(X_{1}^{l-1}), \textit{LN}(X_{2}^{l-1})) + X_{1}^{l-1} \\
		X_{12}^{l} = \textit{MLP}(\textit{LN}(\hat{X}_{12}^{l})) + \hat{X}_{12}^{l}
	\end{gathered}
\end{equation}

WF-MSA is the abbreviation for window fusion multi-headed self-attention. As in the existing approach, our proposed Transformer structure uses self-attention. The self-attention mechanism can be expressed using the following equations:

\begin{equation}
	\begin{gathered}
		\textit{Attention}(Q,K,V)=\textit{Softmax}({\frac{QK^\mathrm{T}}{\sqrt{d} } + W^{B}})\cdot V
	\end{gathered}
\end{equation}

where $Q$, $K$, and $V$ denote the query matrix, the key matrix, and the value matrix, respectively. $W^{B}$ is a parameterized matrix that models relative position information. $d$ is the dimension of the query matrix or the key matrix. Multi-head self-attention is one of the core components of the Transformer. In contrast to self-attention, the multi-head mechanism splits the input into many small parts and concatenates all the outputs to get the final result. As shown in Fig. \ref{FIG:3}(c), given the input modalities $X_{1}$ and $X_{2}$, the formula for our designed WF-MSA can be written as:

\begin{equation}
	\begin{gathered}
		\textit{WF-MSA}(X_{1}, X_{2}) = \textit{Concat}(head_1,...,head_i)W^O \\
		head_i = \textit{Attention}(X_{1}W_i^Q,X_{2}W_i^K,X_{2}W_i^V)
	\end{gathered}
\end{equation}

$W_i^Q$, $W_i^K$, $W_i^V$, and $W^O$ are trainable parameter matrices. In this work, we employ $i = \left\{4, 8, 16, 32\right\}$ parallel attention layers in four stages of multimodal fusion block.

\subsubsection{CNNs-based decoder}
In the decoder component, we mainly use CNNs to form the backbone of our PT-Net. There are two reasons for this. First, the convolution operation in CNNs has a stronger ability to extract local information than the pure Transformer architecture. Second, the pure Transformer structure consumes a lot of computational resources and introduces a large number of model parameters, making the network more difficult to converge. We describe the specific implementation of the model below.

In the encoder, each modality-independent branch outputs feature maps of the same shape. First, we stack the feature maps of different modalities. The stacked feature maps contain information on the three modalities. Since different channels in the feature maps of the three modalities may have different importance, we use an information calibration block to fuse and calibrate the multimodal information. Specifically, we utilize a channel-based squeeze-and-excitation (SE) \cite{hu2018squeeze} attention to reorganize and reallocate information across modalities. The SE attention explicitly models inter-channel dependencies by learning the attention weights of each channel, thus allowing the network to focus more on important channels and improving the performance of the segmentation model. We use the SE attention after each convolutional block of the information calibration module. 

The information calibration module starts with a deconvolution layer. The resolution of the depth dimension of each feature map remains fixed, and the resolution of the other two dimensions becomes twice. After the deconvolution layer, the feature maps are fed into a slightly modified residual block with SE channel attention, while the number of channels becomes one-third of the original. Compared to the regular residual block, we replace the batch normalization with instance normalization and the ReLU with Leaky ReLU.

After the above information calibration module with channel attention, each decoding layer except the bottleneck layer is stacked with the feature map from the previous layer, achieving the skip connection similar to U-Net. Finally, the stacked feature maps go through the modified residual block and change their resolution to twice the original one. 

\subsection{Anatomy-aware loss function}
To handle parotid tumor lesions with different scales and challenging anatomical structures, we propose a new distance-based loss function, anatomy-aware loss, to train the segmentation network.

Suppose the target image is $I$, the ground truth label is $G$, and the model predicted segmentation after the softmax function is $P$. We expect the model prediction to be as close as possible to the ground truth label, so we need to restructure and guide the distribution of the output segmentation probability map of the model prediction. First, we define the activation center coordinates $C$ ($u_c$, $v_c$, $w_c$) in the feature map as follows:

\begin{equation}
	\begin{gathered}
		u_c = \frac{1}{\Omega} \frac{\sum\nolimits_{(u, v, w) \in \Omega} u \cdot I(u, v, w)}{\sum\nolimits_{(u, v, w) \in \Omega} I(u, v, w)}
	\end{gathered}
\end{equation}

where $I(u, v, w)$ is the pixel value in $G$ or $P$ with coordinates $(u, v, w)$, $\Omega$ is the target region, i.e., the set of $G$ or $P$. In this formula, $u_c$ characterizes the central coordinates of the target region in one dimension. However, when the output of the target region is all zero, the denominator of the formula is also zero, thus causing training instability. To overcome this difficulty, we further add the stability constant $\epsilon$:

\begin{equation}
	\begin{gathered}
		u_c = \frac{1}{\Omega} \frac{\sum\nolimits_{(u, v, w) \in \Omega} u \cdot (I(u, v, w) + \epsilon)}{\sum\nolimits_{(u, v, w) \in \Omega} (I(u, v, w) + \epsilon)}
	\end{gathered}
\end{equation}

In our work, $\epsilon$ is set to $10^{-8}$. Similarly, we can obtain the activation center coordinates for the other two dimensions:

\begin{equation}
	\begin{gathered}
		v_c = \frac{1}{\Omega} \frac{\sum\nolimits_{(u, v, w) \in \Omega} v \cdot (I(u, v, w) + \epsilon)}{\sum\nolimits_{(u, v, w) \in \Omega} (I(u, v, w) + \epsilon)}
	\end{gathered}
\end{equation}

\begin{equation}
	\begin{gathered}
		w_c = \frac{1}{\Omega} \frac{\sum\nolimits_{(u, v, w) \in \Omega} w \cdot (I(u, v, w) + \epsilon)}{\sum\nolimits_{(u, v, w) \in \Omega} (I(u, v, w) + \epsilon)}
	\end{gathered}
\end{equation}

Note that we also add the stability constant $\epsilon$ to the numerator. The motivation for adding this constant is that when training with image patches, some of the ground truth labels are all zero. In this case, if the numerator does not have the stability constant, the activation center coordinates are constant at zero. After adding the stability constant, the activation center coordinates become the center of the feature map:

\begin{equation}
	\begin{aligned}
		\lim\limits_{I(u, v, w) \to 0} u_c &= \frac{1}{\Omega} \frac{\sum\nolimits_{(u, v, w) \in \Omega} u \cdot \epsilon}{\sum\nolimits_{(u, v, w) \in \Omega} \epsilon} \\
		&= \frac{1}{\Omega} \sum\nolimits_{i=1}^{\Omega} u \\
		&= \frac{1}{2} u_{max}
	\end{aligned}
\end{equation}

where $u_{max}$ is the maximum value of the coordinates in one dimension of $G$ or $P$. In contrast to the loss function without the stability constant, when the model's prediction is segmented mistakenly in a blank background, the activation center coordinates of the model prediction and the ground truth deviate on a large scale. Finally, the loss function can successfully compute the loss and backpropagate the gradient. Thus, the presence of the stability constant explicitly suppresses the problem that the model still outputs the segmentation incorrectly when the patch-based ground truth is all the background.

We compute the activation centers for the two matrices of the ground truth $G$ and the model prediction $P$, respectively. Finally, we obtain two center coordinates $C_G$ ($u_g$, $v_g$, $w_g$) and $C_P$ ($u_p$, $v_p$, $w_p$). We want to avoid anatomical structures in ground truth labels where incorrect inference occurs at a large distance or tumors are missing. Therefore, we calculate the Euclidean distance for those two activation center coordinates as follows:

\begin{equation}
	\begin{gathered}
		\mathcal{L}_{A} = \lvert u_g-u_p \rvert^{\beta}d_u + \lvert v_g-v_p \rvert^{\beta}d_v + \lvert w_g-w_p \rvert^{\beta}d_w
	\end{gathered}
\end{equation}

$\mathcal{L}_{A}$ is our proposed anatomy-aware loss. $\beta$ is the distance penalty factor to penalize inference errors at large distances. The penalty factor promotes the model to produce awareness of anatomical structures. $D$ = ($d_u$, $d_v$, $d_w$) is the spacing of the 3D MRI, which is used to map the image pixel to the physical distance.

\section{Experiments and results}
\subsection{Experimental setting}
\subsubsection{Data and implementation details}
In this work, we used MRI datasets of parotid tumor patients to validate our automatic segmentation framework. Patients who underwent MRI examinations at the China Medical University School and Hospital of Stomatology from 2013 to 2020 were included in the experimental study. Following initial quality assessment and selection based on images, we obtained 187 MRIs of parotid tumor patients. Each patient's MRI was available in three modalities, T1, T2, and STIR images. The ground truth was annotated by well-trained medical students and was checked and reviewed by experts to ensure the accuracy of the labels. All patients were aged between 7 and 86 years, with a male to female ratio of 1.3 to 1. The Ethics Committee has approved the use of MRI of patients for this work.

The intra-slice pixel spacing ranged from 0.28 mm to 0.43 mm, with a slice size of 512 × 512 or 640 × 640. There was not a wide range of inter-slice spacings of 3.85 mm to 6.50 mm, with a mean value of 4.44 mm. With a mean value of 17.23, the slice number ranged from 13 to 22. In our experiments, we used the patch-based training strategy. Precisely, we randomly cropped patch sizes of 8 × 320 × 320 from the images and fed them into the model. At least half of the image patches in each batch should contain foreground regions to improve training stability. 

Our automatic segmentation framework experiments were implemented using PyTorch with an Nvidia RTX 3060 12G GPU. To fairly compare the effectiveness of the models, we adopted an experimental setup similar to that of the popular medical image segmentation framework nnU-Net \cite{isensee2021nnu}. For non-nnUNet experiments, the batch size was set to 2. All experiments were trained for 200 epochs, containing 250 mini-batches in each epoch. We trained the model using the stochastic gradient descent (SGD) optimizer with Nesterov momentum of 0.95. During training, the initial learning rate was set at 0.01 and decayed after each epoch. In order to achieve the final segmentation results, the data was inferred using the sliding window technology after the training process had been completed. It is worth noting that parotid tumors represent a very small foreground in the images. Therefore, five-fold cross-validation was performed for each experiment to reduce the risk of model overfitting, and paired t-test was used to highlight significance.

In the loss function experiments, we did not directly use the anatomy-aware loss in Eq. 13. Distance-based loss functions used alone may cause training instability and model collapse \cite{kervadec2019boundary}. The distance loss is used with Dice loss to enable fast convergence and performance improvement. Therefore, distance-based benchmark methods are paired with Dice loss in our loss experiments. We also combined anatomy-aware loss with the Dice loss to fairly compare the performance of the proposed method with other distance-based losses. In particular, our training strategy follows the Hausdorff Distance (HD) loss \cite{karimi2019reducing}:

\begin{equation}
	\begin{gathered}
		\mathcal{L} = \mathcal{L}_{Distance} + \lambda \mathcal{L}_{Dice}
	\end{gathered}
\end{equation}

$\mathcal{L}_{Distance}$ represents the distance-based loss, $\mathcal{L}_{Dice}$ represents the dice loss, and $\lambda$ denotes the weighting factor. Following HD loss, we can choose $\lambda$ so that the distance-based loss term and the Dice loss term have equal weights. After each training epoch, we computed the distance-based loss and the Dice loss on the training data, and for the next epoch, $\lambda$ should be set as the ratio of the mean of the distance-based loss to the mean of the Dice loss.

\subsubsection{Evaluation metrics}
To further explore the impact of tumors of different size on the automatic segmentation model, our work divided the dataset into three groups. Group A contains 53 images of parotid tumors over 10 cubic centimeters (CC), group B includes 59 images of parotid tumors between 4 and 10 CC, and group C contains 75 images of parotid tumors less than 4 CC. 

For quantitative evaluation, we measured precision, recall, Dice coefficient, Average Surface Distance (ASD), and the 95th percentile of the Hausdorff Distance (HD$_{95}$) between segmentation results and ground truth.

\begin{equation}
	\begin{gathered}
		Precision = \frac{TP}{TP+FP} 
	\end{gathered}
\end{equation}

\begin{equation}
	\begin{gathered}
		Recall = \frac{TP}{TP+FN}
	\end{gathered}
\end{equation}

\begin{equation}
	\begin{gathered}
		Dice = \frac{2 \times TP}{2 \times TP + FN + FP}
	\end{gathered}
\end{equation}

where TP, TN, FP, and FN are true positive, true negative, false positive, and false negative, respectively.

\begin{equation}
	\begin{gathered}
		\textit{HD} = max\left(\mathop{max}\limits_{p \in P} d(p, G), \mathop{max}\limits_{g \in G} d(g, P)\right)
	\end{gathered}
\end{equation}

\begin{equation}
	\begin{gathered}
		\textit{ASD} = \frac{1}{\lvert P \rvert} \sum\limits_{p \in P}d(p, G)
	\end{gathered}
\end{equation}

$P$ and $G$ represent the set of surface points of the prediction result and the ground truth segmentation, respectively. $d(p, G)$ is the shortest distance between a point $p \in P$ and all the points in $G$.

\subsection{Ablation studies}
In this section, we conducted the ablation study of the model and reported the experimental results. First, to demonstrate the robustness of the loss function, we performed experiments on the parameters of our anatomy-aware loss. Second, we investigated the validity of the framework by examining two components of the automatic segmentation model. Finally, we conducted ablation experiments on different input image modalities to investigate the effect of multimodal fusion.

\subsubsection{Comparison of different parameters in the anatomy-aware loss}

\begin{table}[width=.9\linewidth,cols=4,pos=h]
	\caption{Quantitative comparison of our proposed anatomy-aware loss function from different training parameters. The models are all trained on PT-Net.}
	\label{tbl2}
	\begin{tabular*}{\tblwidth}{@{} LLLLLL@{} }
		\toprule
		Params & Precision (\%) & Recall (\%) & Dice (\%)   & HD95 (mm)   & ASD (mm)   \\ \midrule
		$\beta=0.5$  & 84.19±17.14    & 81.11±19.36 & 80.62±17.31 & 13.50±30.14 & 2.95±10.61 \\
		$\beta=1.0$  & 85.47±15.36    & 81.46±17.82 & 82.10±14.91 & 8.71±22.09  & 1.83±6.67  \\
		$\beta=1.5$  & 86.17±12.09    & 82.57±17.40 & 82.57±14.63 & 8.05±21.25  & 1.22±2.70  \\
		$\beta=2.0$  & 85.17±16.47    & 80.35±18.89 & 81.29±16.00 & 10.30±25.93 & 2.26±12.70 \\
		$\beta=2.5$  & 83.51±20.45    & 77.95±22.83 & 78.96±20.60 & 11.54±27.54 & 3.25±16.18 \\ \bottomrule
	\end{tabular*}
\end{table}

In this section, we explored the effect of different penalty factors in the anatomy-aware loss on the model performance to verify the robustness of the proposed loss function. Table \ref{tbl2} reports the experimental results. From the table, it can be found that the model has the best performance in all aspects when $\beta$ = 1.5. When $\beta$ = 1.0, the model achieves the second-best results with an precision of 85.47 $\pm$ 15.36\%, a recall of 81.46 $\pm$ 17.82\%, a Dice of 82.10 $\pm$ 14.91\%, an HD$_{95}$ of 8.71 $\pm$ 22.09 mm, and an ASD of 1.83 $\pm$ 6.67 mm. However, the model performance decreases when $\beta$ is too large. At $\beta$ = 2.5, the performance of our anatomy-aware loss was worse than the combined loss function of Dice and CE. We set $\beta$ = 1.5 in the following experiments.

\begin{table}[width=.9\linewidth,cols=4,pos=h]
	\caption{Quantitative evaluation of different networks for parotid tumor segmentation. PT-Net: The proposed network for parotid tumor segmentation. SE: SE attention module. MF: multimodal fusion block. The best result of each metric is highlighted in bolded, and the second best result is underlined. All results include the mean and standard deviation of the cross-validation trials.}
	\label{tbl3}
	\begin{tabular*}{\tblwidth}{@{} LLLLLL@{} }
		\toprule
		Network         & Precision (\%)       & Recall (\%)          & Dice (\%)            & HD95 (mm)            & ASD (mm)           \\ \midrule
		nnUNet 2D       & 75.89±27.99          & 73.54±28.76          & 72.88±26.91          & 17.54±33.71          & 8.54±25.07         \\
		nnUNet          & 79.88±22.84          & 80.17±21.78          & 77.98±20.96          & 19.65±37.67          & 6.50±17.45         \\
		Attention UNet  & 76.74±27.85          & 75.69±28.67          & 75.19±27.02          & 18.73±37.51          & 9.83±28.76         \\
		ResUNet         & 80.17±24.06          & 74.39±25.55          & 74.91±23.60          & 16.58±32.38          & 6.51±18.73         \\
		SegResNet       & 76.61±27.67          & 75.07±27.10          & 74.30±25.88          & 16.39±32.35          & 7.53±23.13         \\
		UNETR           & 79.76±21.25          & 77.22±25.50          & 75.81±23.30          & 17.57±34.54          & 5.93±17.91         \\
		Swin UNETR      & 79.38±21.10          & {\ul 79.92±22.64}    & 77.00±21.89          & 21.37±38.89          & 6.62±15.86         \\
		Swin UNet       & 81.60±17.10           & 77.42±24.06          & 76.43±22.15          & 15.61±32.27          & 4.37±11.85         \\
		CoTr            & 80.41±21.61          & 79.54±23.04          & 78.34±20.82          & 12.56±29.22          & 4.37±18.32         \\
		PT-Net          & \textbf{86.06±14.70} & 79.41±19.62          & \textbf{80.87±16.79} & \textbf{10.22±23.98} & \textbf{2.42±7.07} \\
		PT-Net (w/o SE) & 80.95±19.25          & \textbf{82.02±20.24} & {\ul 79.65±18.47}    & {\ul 12.28±28.92}    & 4.42±17.14         \\
		PT-Net (w/o MF) & {\ul 82.61±19.03}    & 78.67±22.37          & 78.12±21.00          & 13.98±29.96          & {\ul 4.02±11.07}   \\ \bottomrule
	\end{tabular*}
\end{table}

\subsubsection{Comparison of different network architectures}
To investigate the effect of network architecture on our automated parotid tumor segmentation, we compared PT-Net with (1) PT-Net w/o SE, where PT-Net has an information calibration module without the SE attention module, (2) PT-Net w/o MF, where PT-Net uses a standard modality-independent Transformer block instead of the multimodal fusion block, (3) 2D nnU-Net, (4) 3D nnU-Net, (5) Attention U-Net \cite{oktay2018attention}, (6) ResUNet \cite{xiao2018weighted}, (7) SegResNet \cite{myronenko20183d}, (8) UNETR \cite{hatamizadeh2022unetr}, (9) Swin UNETR \cite{hatamizadeh2022swin}, (10) Swin U-Net \cite{cao2021swin}, and (11) CoTr \cite{xie2021cotr}. Table \ref{tbl3} reports the quantitative comparison between the different network architectures. Among all segmentation networks, the proposed PT-Net achieved the highest precision, Dice coefficient, HD$_{95}$, and ASD. The PT-Net without SE achieved slightly better recall evaluation results. It can be seen that the combination of the SE attention module and multimodal fusion block is superior to using a single module. Our PT-Net obtained the highest average Dice of 80.87\%. This is a significant improvement compared to the best baseline method CoTr with 78.34\%.

\begin{figure}
	\centering
	\includegraphics[scale=.6]{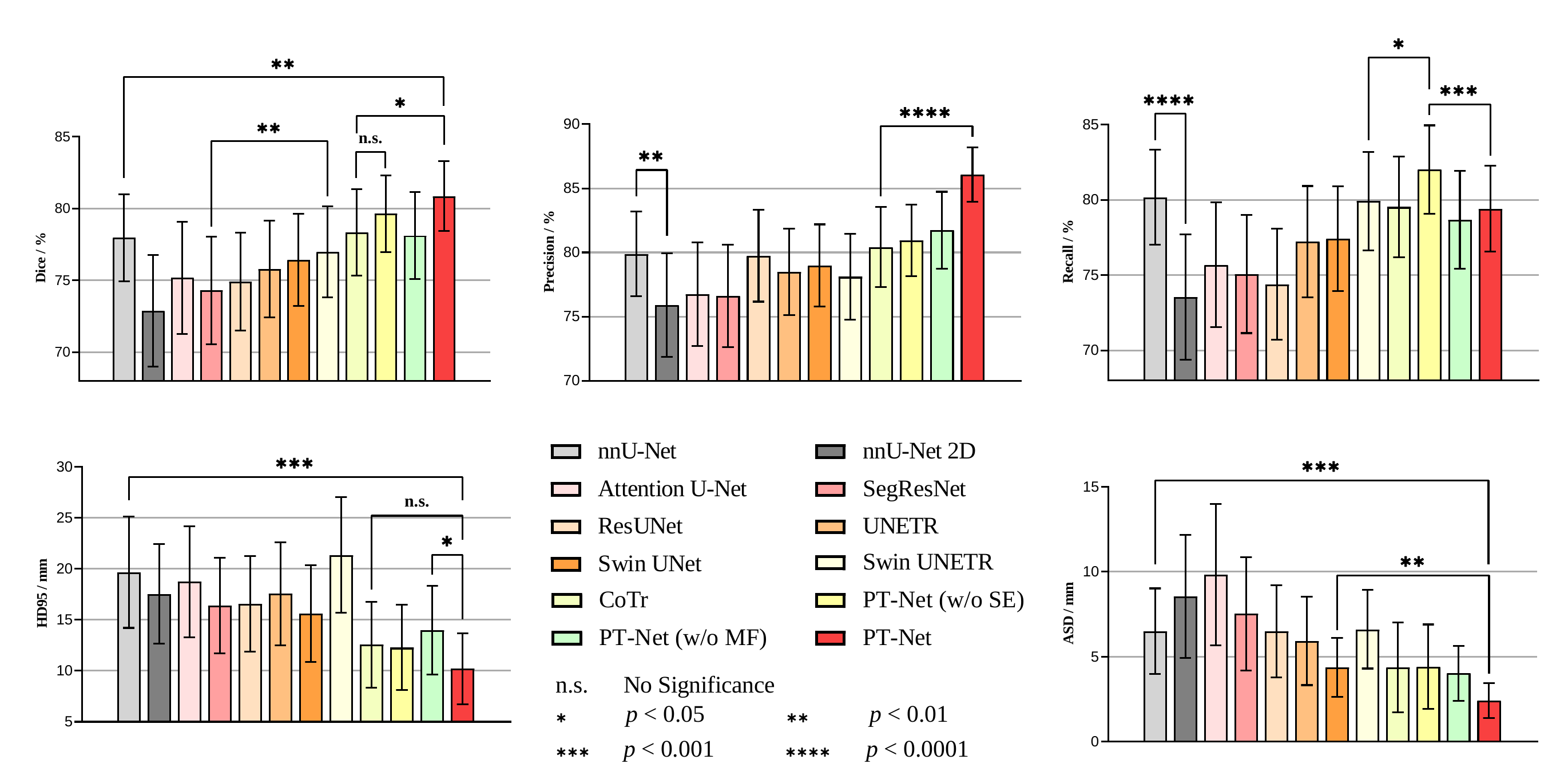}
	\caption{Results of paired t-test for different network architectures in parotid tumor segmentation.}
	\label{FIG:4}
\end{figure}

\begin{figure}
	\centering
	\includegraphics[scale=.43]{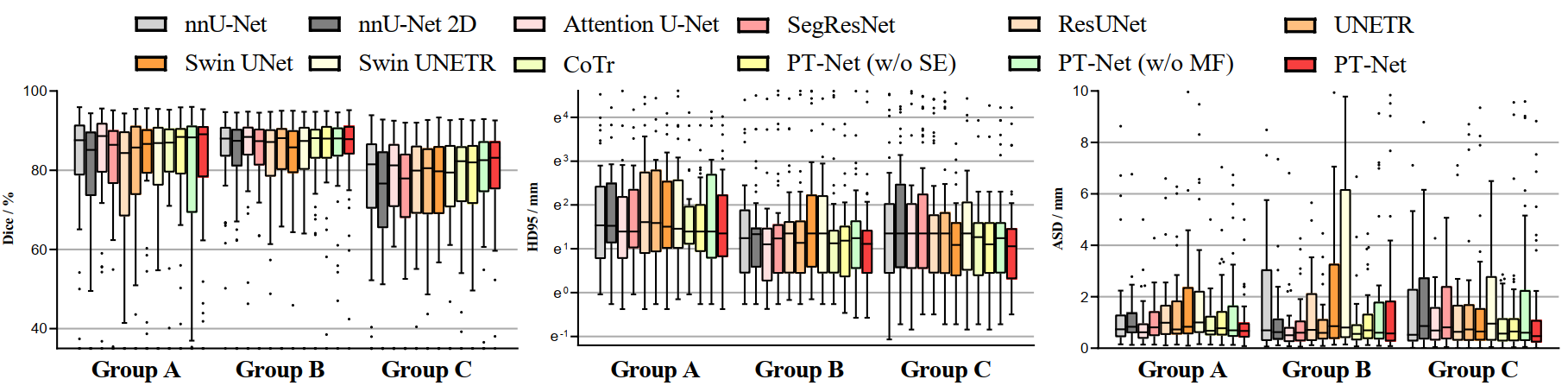}
	\caption{Group-wise comparison of different network architectures for parotid tumor segmentation. Group A: tumors larger than 10 CC. Group B: tumors between 4 and 10 CC. Group C: tumors smaller than 4 CC.}
	\label{FIG:5}
\end{figure}

\begin{figure}
	\centering
	\includegraphics[scale=.8]{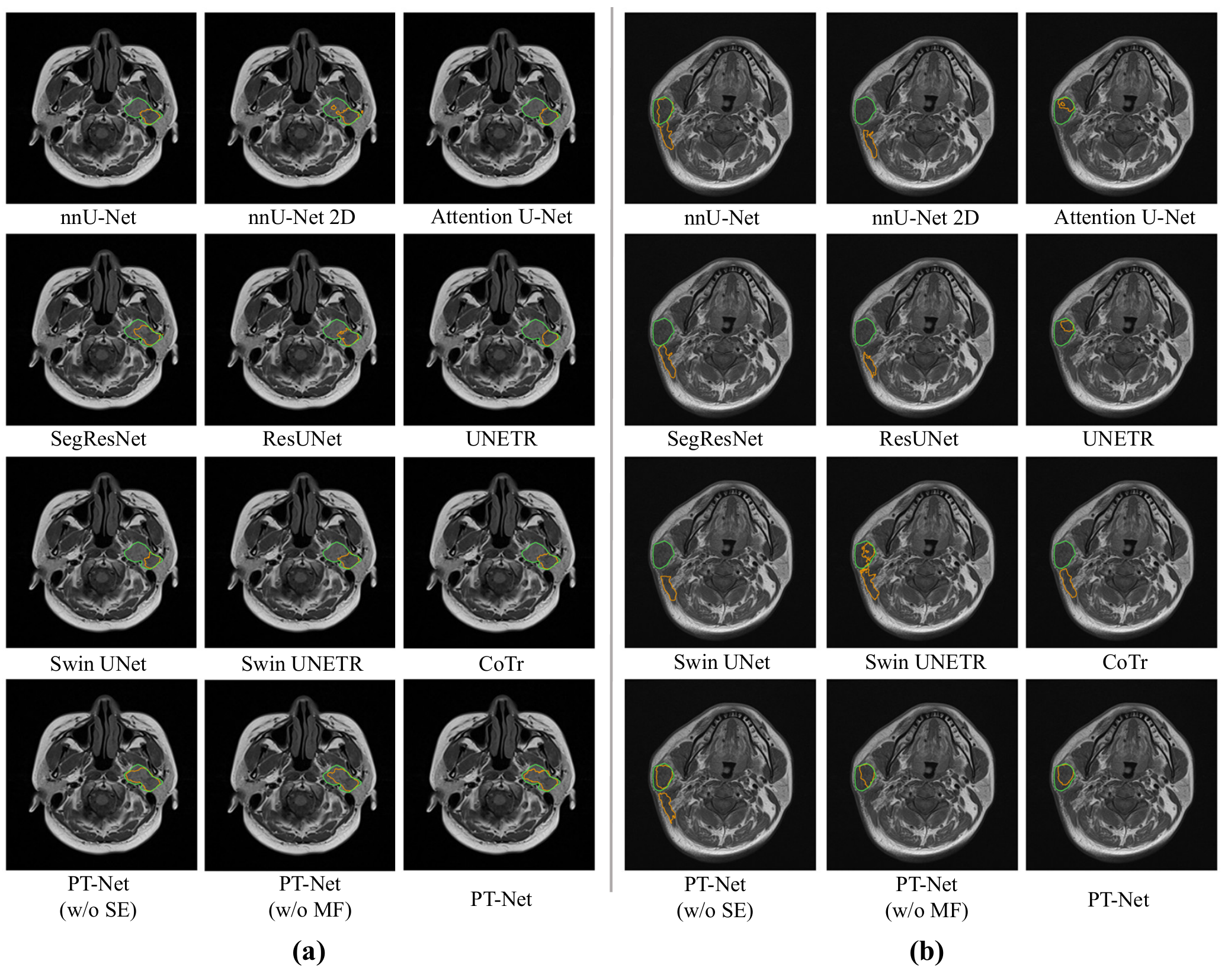}
	\caption{Qualitative comparison of the different networks used for parotid tumor segmentation. (a) and (b) are from two different individuals. The orange contours represent the model prediction segmentation results. The green contours represent the results of the ground truth segmentation.}
	\label{FIG:6}
\end{figure}

Fig. \ref{FIG:4} shows paired t-test results of different compared methods. From the observation, we can obtain that PT-Net is significant compared to the best baseline method in terms of Dice coefficient, precision, and ASD. Although there is no significant difference between PT-Net and CoTr in HD$_{95}$, there is still a performance gap of 2.34 mm. In addition, we can see that 2D nnU-Net performs much lower than 3D nnU-Net and other methods. It shows that the 2D network does not effectively capture the spatial relationship and semantic information of the tumor. Fig. \ref{FIG:5} shows the quantitative comparison of these networks in three groups of images at different lesion scales. The results show that our proposed PT-Net achieves the best performance in different scales of tumors and has good robustness. Fig. \ref{FIG:6} visually compares the segmentation results of PT-Net and all baselines. These examples cover lesions of different scales. These visual comparisons show that PT-Net outperforms other networks when segmenting challenging tumors of varying scales.

\begin{table}[width=.9\linewidth,cols=4,pos=h]
	\caption{Quantitative evaluation of different loss functions for parotid tumor segmentation based on our proposed PT-Net. AMA: anatomy-aware loss function.}
	\label{tbl4}
	\begin{tabular*}{\tblwidth}{@{} LLLLLL@{} }
		\toprule
		Loss       & Precision (\%)       & Recall (\%)         & Dice (\%)            & HD95 (mm)           & ASD (mm)           \\ \midrule
		CE         & 85.31±15.24          & 79.05±22.11         & 79.65±19.72          & {\ul 10.07±26.23}   & 2.64±13.84         \\
		Dice       & 84.14±17.41          & {\ul 80.74±18.56}   & 80.68±17.14          & 13.46±31.01         & 4.17±13.06         \\
		Dice + CE    & {\ul 86.06±14.70}    & 79.41±19.62         & {\ul 80.87±16.79}    & 10.22±23.98         & {\ul 2.42±7.07}    \\
		Focal      & 75.29±28.20          & 72.37±28.88         & 71.18±27.42          & 20.66±37.28         & 9.31±23.56         \\
		Dice + Focal & 84.09±16.89          & 79.30±21.90         & 79.55±19.64          & 12.54±28.90         & 3.06±12.25         \\
		Dice + BD    & 79.27±23.73          & 75.65±23.07         & 75.55±22.21          & 25.26±43.03         & 9.40±20.38          \\
		Dice + HD    & 80.71±21.71          & 78.68±22.98         & 77.54±21.32          & 17.15±34.70         & 6.24±18.87         \\
		Dice + AMA (ours)   & \textbf{86.17±12.09} & \textbf{82.57±17.40} & \textbf{82.57±14.63} & \textbf{8.05±21.25} & \textbf{1.22±2.70} \\ \bottomrule
	\end{tabular*}
\end{table}

\begin{figure}
	\centering
	\includegraphics[scale=.6]{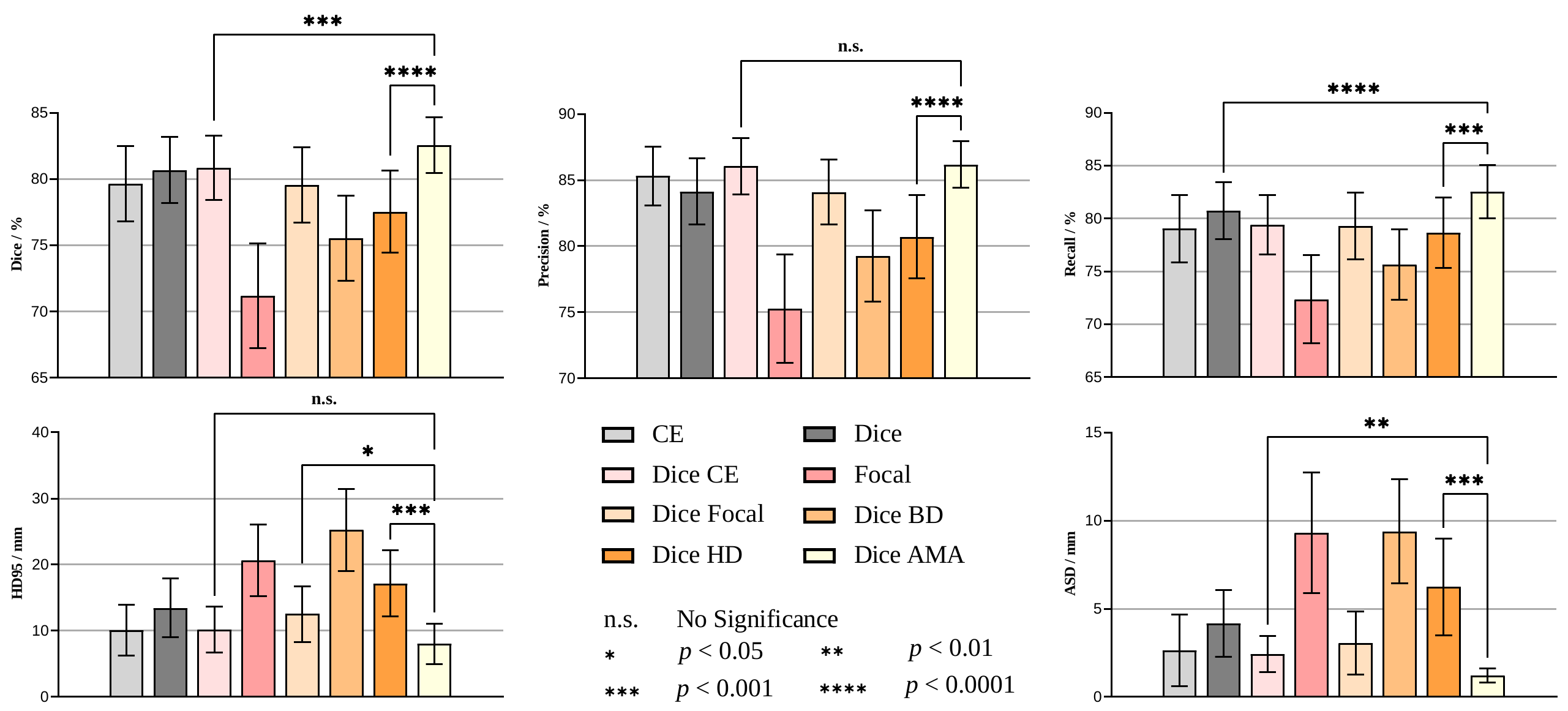}
	\caption{Results of paired t-test for different loss functions in parotid tumor segmentation.}
	\label{FIG:7}
\end{figure}

\begin{figure}
	\centering
	\includegraphics[scale=.43]{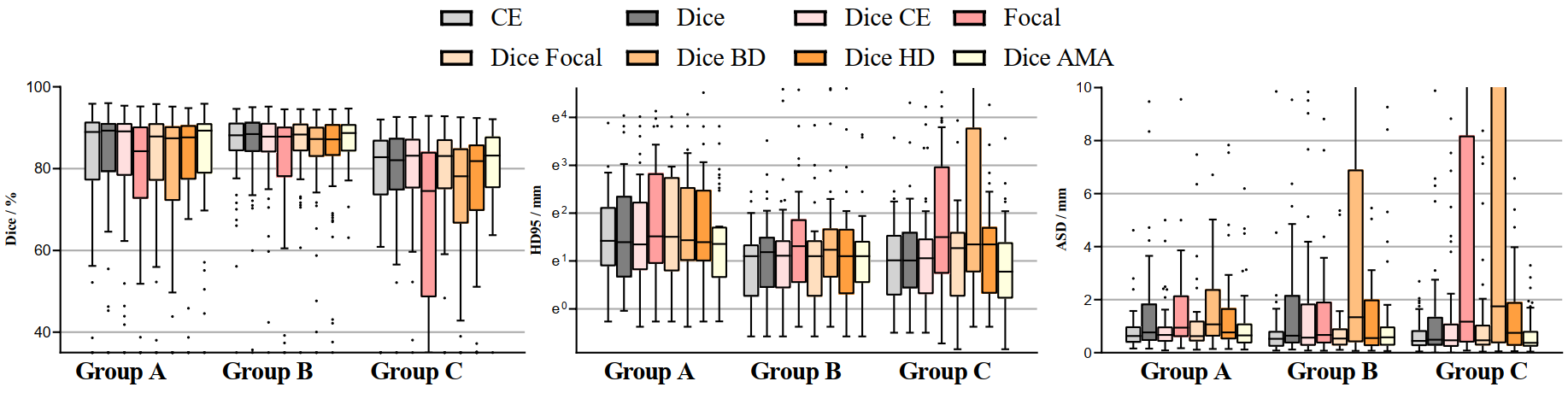}
	\caption{Group-wise comparison of different loss functions for parotid tumor segmentation.}
	\label{FIG:8}
\end{figure}

\begin{figure}
	\centering
	\includegraphics[scale=.65]{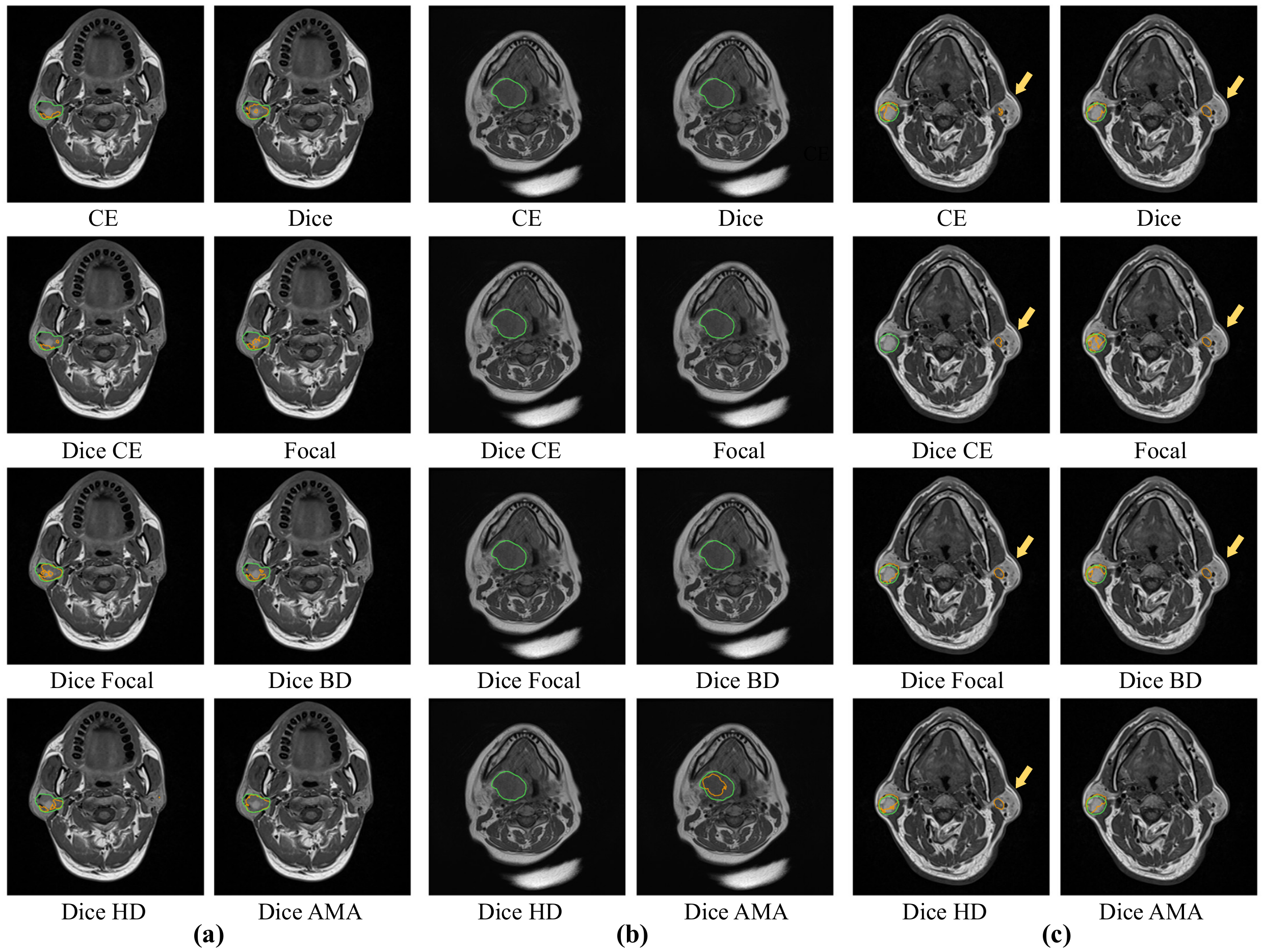}
	\caption{Qualitative comparison of the different loss functions used for parotid tumor segmentation. (a), (b) and (c) are from three different individuals. Yellow arrows represent some incorrectly segmented regions.}
	\label{FIG:9}
\end{figure}

\subsubsection{Comparison of different training loss functions}
To further explore the effect of anatomy-aware loss function on the performance of parotid tumor segmentation, we compare the proposed loss with several state-of-the-art methods. Most loss functions combine multiple losses to ensure good training stability and performance. Among all compared methods, there are three single losses, which are (1) CE loss, (2) Focal CE loss \cite{lin2017focal}, and (3) Dice loss. There are four compound losses, which are (4) Dice CE loss, (5) Focal Dice loss \cite{zhu2019anatomynet}, (6) Dice HD loss \cite{karimi2019reducing}, and (7) Dice boundary (BD) loss \cite{kervadec2019boundary}.

Table \ref{tbl4} reports the performance of the deep model with different loss functions for automatic segmentation of parotid tumors. The compound of Dice and CE loss consists of good performance in baseline loss functions. The results show that the performance of Focal CE loss is relatively low, probably because it focuses too much on the hard samples. The two distance-based baseline loss functions, i.e., HD loss and BD loss, perform poorly in the experiments, having lower Dice coefficients and higher HD$_{95}$. This is probably the overfitting of the model to the disturbed anatomy. In addition, these two methods are computationally slow and difficult to compete with other methods. The anatomy-aware loss proposed in this paper reaches the most advanced performance among all methods, with Dice at 82.57 $\pm$ 14.63\% and HD$_{95}$ at 8.05 $\pm$ 21.25 mm. In particular, the value of our method in the ASD decreases by 49\% compared to the second-best model. This shows the effectiveness of the proposed method in encouraging the model to focus on the ground truth region. Also, the training speed of the anatomy-aware loss is almost the same as the other baseline loss functions, demonstrating a good balance of speed and accuracy.

Fig. \ref{FIG:7} shows the results of paired t-tests with different loss functions. As can be seen from the figure, our proposed method brings significant improvements. Notably, the anatomy-aware loss dramatically outperforms the other two distance-based loss functions in each metric (p<0.001). There is no statistical significance between the anatomical perception loss and the subsequent best Dice CE loss in the HD$_{95}$ and accuracy. However, there is a large performance gap in the other metrics. A quantitative comparison of these networks for each group of images is given in Fig. \ref{FIG:8}. It can be observed from the figure that our proposed anatomy-aware loss can bring performance gains for tumors at different scales. Fig. \ref{FIG:9} shows the visual comparison of tumor segmentation for the models trained with different loss functions. It can be observed that our anatomy-aware loss performs better than the others, which indicates the advantage of the proposed loss in excluding the anatomical regions.

\subsubsection{Comparison of different input image modalities}

\begin{table}[width=.9\linewidth,cols=4,pos=h]
	\caption{Quantitative comparison of parotid tumor segmentation from different input images based on proposed PT-Net.}
	\label{tbl5}
	\begin{tabular*}{\tblwidth}{@{} LLLLLL@{} }
		\toprule
		Modality     & Precision (\%)       & Recall (\%)          & Dice (\%)            & HD95 (mm)            & ASD (mm)           \\ \midrule
		T1           & 80.60±21.42          & 77.03±24.03          & 76.28±22.03          & 15.81±32.28          & 4.93±14.59         \\
		STIR         & 81.76±21.68          & 75.11±23.84          & 76.19±21.80          & 14.97±31.74          & 5.56±19.87         \\
		T1, T2       & 84.63±17.12          & 77.63±23.09          & 78.53±20.60          & 11.16±26.07          & 2.61±7.01    \\
		T1, STIR     & 85.03±16.86    & 78.73±20.78    & 80.03±18.22    & 10.06±25.54 & 3.00±12.79         \\
		T1, T2, STIR & 86.06±14.70 & 79.41±19.62 & 80.87±16.79 & 10.22±23.98    & 2.42±7.07 \\ \bottomrule
	\end{tabular*}
\end{table}

To investigate the effect of different input modalities on the automatic segmentation of parotid tumors, we compared the default input modalities (T1, T2, and STIR MRI) with (1) T1 MRI only, (2) STIR MRI only, (3) T1 and T2 MRI, and (4) T1 and STIR MRI. However, due to the multimodal fusion block, PT-Net does not directly support the input of images from one or two modalities. To solve this problem, we slightly modified the multimodal fusion block. On the one hand, when there is only one modality input, the multimodal fusion block degenerates to the window block, which is the same as Swin Transformer. On the other hand, when two input modalities exist, only one window fusion block is kept in the multimodal fusion block to compute the feature maps of the two inputs. Table \ref{tbl5} reports the quantitative results using different input images. From the experimental results, it can be observed that the performance of the model keeps improving by gradually fusing multiple image modalities. Notably, the performance of T1 and STIR MRI is considerably higher than that of T1 and T2 MRI when there are two modality images input and only slightly lower than the default setting of three modality images. The higher similarity and mutual information of T1 and T2 images compared to STIR MRI is possibly responsible for the difference in model performance.

\section{Discussion}
The parotid gland and its surroundings are complex and have a large number of anatomical structures. These structures have similar signal intensities to the tumor in MRI, posing a significant challenge to tumor segmentation. To address this problem, we propose an anatomy-aware framework for automatic parotid tumor segmentation in this paper.

Considering the difficulty of directly distinguishing anatomical structures from tumor regions, learning prior knowledge from multimodal images is possibly a practical solution. We propose PT-Net, a novel parotid tumor segmentation network. The architecture of PT-Net hybridizes CNNs and the Transformer to better model local lesion features and long-range dependence information. We note that existing work, such as CoTr \cite{xie2021cotr}, UNETR \cite{hatamizadeh2022unetr}, and Swin UNETR \cite{hatamizadeh2022swin}, also propose hybrid architectures, but these networks have critical differences from ours. First, we use the multimodal fusion block. The multimodal fusion block fuses the feature maps of multiple modality-independent encoder outputs from coarse to fine. The proposed window fusion multi-head attention facilitates the interaction and flow of cross-modal information through self-attention. Second, we use the information calibration module in the decoder to enhance the network's ability to handle multimodal information. In particular, the channel-based SE attention calibrates and reweights information from the feature maps of multiple modalities. Experiments demonstrate that our fusion technology improves the performance of automatic segmentation of parotid tumors.

In Fig. \ref{FIG:6}, the visualization results exhibit the advantages of our proposed PT-Net in parotid tumor segmentation. As seen in Fig. \ref{FIG:6}(a), PT-Net still performs well in segmenting the relatively rare deep lobe tumors. However, other networks cannot distinguish tumors in the deep part of the parotid gland well. The comparison results in Fig. \ref{FIG:6}(b) show that the proposed PT-Net can achieve more accurate segmentation than the baseline method when the signal intensities of the muscle tissue and the lesion are similar.

Furthermore, we present a novel distance-based loss function. The anatomy-aware loss indirectly facilitates the model's self-awareness of anatomical structures, thus reducing incorrect segmentation for disturbing structures. Our loss explicitly penalizes incorrect predictions far from the ground truth labels to force the model to focus on the difference between anatomical structures and tumor regions. Compared to the current distance-based loss functions \cite{kervadec2019boundary,karimi2019reducing}, the proposed method has three significant benefits in parotid gland segmentation. First, the anatomy-aware loss does not require the computation of distance maps in training, making the calculation much faster. Second, in parotid gland MRI, the tumor region is often small, while the anatomical structures are far away from the tumor region. In training, the value of the distance-based loss function fluctuates considerably. However, our loss function has a relatively small maximum value. Therefore, the anatomy-aware loss is more stable in training than HD loss and BD loss. Third, the distance-based loss has the smallest value when the model predicts all zero results, which causes performance degradation. In contrast, our proposed loss can still calculate when the model prediction is all zero and guide the model to correct the interference region.

Fig. \ref{FIG:9} illustrates the visual comparison of anatomy-aware loss and other loss functions. There is a high degree of inconsistency in the signal intensity in the tumor region in Fig. \ref{FIG:9}(a) and Fig. \ref{FIG:9}(b), and the tumor boundary is ambiguous. The baseline method has big mistakes at the boundary. On the contrary, our method has the best segmentation results. The segmentation in Fig. \ref{FIG:9}(c) can clearly demonstrate the great advantage of our anatomy-aware loss in perceiving anatomical structures. All other methods incorrectly segment the blood vessels on the right side of the image as the tumor region. However, our method avoids this interference region well, further evidencing the robustness and effectiveness of the anatomy-aware loss in distinguishing anatomical structures.

Although the automatic segmentation framework proposed in this paper achieves satisfactory performance, two additional aspects can be further improved. First, compared with other feature-level fusion approaches, our proposed PT-Net dramatically reduces the computational cost by not stacking feature maps in the encoder. However, it still has three modality-independent encoders that introduce a large number of training parameters. Therefore, we will explore the design of a more efficient fusion approach that reduces the redundant parameters. Second, our method does not precisely model the positional relationship between parotid tumors and individual anatomical structures. Future research plans include using prior knowledge of parotid tumors to design more efficient loss functions to improve the model's performance.

\section{Conclusion}
This paper presents a novel Transformer-based fusion network structure and an anatomy-aware loss function for automatic segmentation of parotid tumors from multimodal MRI. To deal with complex tumors and irregular anatomical structures in MRI volumes, we propose PT-Net, which extracts cross-modality information and long-range dependencies with multimodal fusion block. To leverage prior knowledge of anatomical structures, we propose anatomy-aware loss, a distance-based loss function, where the activation center coordinate is introduced to distinguish interfering structures. Experimental results with parotid gland MRI scans showed that our PT-Net outperformed existing CNNs, Transformer, and hybrid architecture networks for parotid tumor segmentation. Furthermore, anatomy-aware loss could better leverage information about anatomical structures than state-of-the-art loss functions. Our methods can be extended to deal with other lesions and diseases in the future.

\bibliographystyle{cas-model2-names}

\bibliography{cas-refs}

\end{document}